# The condition for the conservation of momentum at the interface under phase transitions of solutions.

Alex Guskov

*Institute of Solid State Physics RAS, Chernogolovka, Moscow region, 142432, Russia.*

*guskov@issp.ac.ru*

**Abstract.** The present work considers a change in the momentum under the transfer of a solution through the interface. It is shown that pressure related to the partial volumes of components arises in a solution under diffusion. As a result, the distribution of the concentration of solution components differs qualitatively from the known solutions. In contrast to the description of one-dimensional interphase mass transfer using the convective diffusion problem, the proposed model provides the stationary exponential distribution of components in both phases. The model describes component segregation by the interface, observed in the experiments.

**Keywords:** phase transition, solution, diffusion, non-equilibrium thermodynamics.

## 1. Introduction.

The study of component diffusion during phase transitions is widely represented in scientific and technical literature [1,2]. One of the tools of these investigations is the convection-diffusion equation. It is used, for example, when studying phase transitions of condensation [1,3], crystallization [3,4], in chemical engineering [5], processes of separation of mixture components [6], polymer production [7], in biology [8]. However, when describing some problems of heterogeneous systems, the use of well-known solutions of the convection mass transfer equation provokes well-founded objections. For example, when describing mass flows through the interfaces [3,6,8], the convection-diffusion equation does not provide segregation of solution components observed in experiments. To describe such experiments, one has to use the Darcy equation, which describes the passage of solutions through porous media [8], or to assume that segregation, for example, under zone melting is related to the hydrodynamic flows of a solution in front of the interface [3,6]. Today, the convection-diffusion equation is used as the model of one-dimensional stationary convection diffusion, which is written in terms of coordinates fixed relative to the interface [3]

$$D_i \frac{\partial^2 c_{Bi}(z)}{\partial z^2} - w \frac{\partial c_{Bi}(z)}{\partial z} = 0. \quad (1)$$

Here $c_{Bi}(z)$ is the concentration distribution of the component B, z is the spatial variable, $D_i$ is the diffusion coefficient. It is assumed that a solution moves in the positive z direction at the specified constant velocity w. It is also assumed that the system under study is heterogeneous, the index i denotes the liquid $i = liq$ or solid $i = sol$ phase. The crystallization of a liquid solution is considered in the present work as an example of a heterogeneous system. The general solution to Eq. (1) has the form

$$c_{Bi}(z) = A_{1i} \exp\left(\frac{wz}{D_i}\right) + A_{2i},$$

where $A_{1i}$ and $A_{2i}$ are the integration constants. To obtain a partial solution to Eq. (1), two boundary conditions are set. The concentration at an infinite point is maintained by the constant

$$c_{Bliq}(\infty) = C_{inf}. \quad (2)$$

The condition for the conservation of a component mass flow at the interface is

$$D_{liq} \frac{dc_{Bliq}(z)}{dz}\bigg|_{z=0} - D_{sol} \frac{dc_{Bsol}(z)}{dz}\bigg|_{z=0} = \left(c_{Bliq}(0) - c_{Bsol}(0)\right)w. \quad (3)$$

The problem provides the concentration distribution

$$c_{Bliq}(z) = C_{inf}\left(1 + \left(\frac{1}{k_e} - 1\right)\exp\left(\frac{wz}{D_{liq}}\right)\right), \quad c_{sol}(z) = C_{inf}, \quad (4)$$

in the liquid and solid phases, respectively. Here $k_e$ is the equilibrium coefficient of the segregation $k_e = c_{sol}(0)/c_{liq}(0)$. According to these solutions, the concentration in the solid phase is constant and does not depend on the velocity of solution transfer. It is equal to the specified concentration of a liquid solution $C_\infty$ (2). The concentration of a liquid solution at the interface does not depend on solution velocity. At any stationary velocity of a solution (including arbitrarily small), the concentration at the interface is $c_{liq}(0) = k_e / c_{sol}(0)$.

The proposed work was motivated by the experiments described in [9,10]. These experiments investigated the solidification of a layer of an aqueous dye solution. The solidification was accompanied by the stationary almost complete displacement of the dye from water without a marked hydrodynamic flow. Problem (1) – (4) does not contain such solutions. If Eq. (1) were a second-order complete linear equation and its solution contained two exponents, the second exponent would describe the concentration distribution in the solid phase, and the problem for a heterogeneous system would have a complete solution. Formally, one can obtain such a solution if one considers a system of equations consisting of Fick's first law and the equation of the conservation of a component mass flow

$$c_{Bi}(z)(w_{Bi}(z) - w) = -D_i \frac{dc_{Bi}(z)}{dz}, \quad (5)$$

$$\frac{d}{dz}(c_{Bi}(z)w_{Bi}(z)) = 0. \quad (6)$$

Let us make transformations of this system. We substitute the expression of a component mass flow, obtained from Eq. (5), into conservation equation (6). As a result, we obtain Eq. (1). Then, we express the concentration gradient from Eq. (5) and substitute it into Eq. (1). As a result of this transformation, the linear equation is obtained

$$D_i \frac{\partial^2 c_{Bi}(z)}{\partial z^2} + \frac{1}{D_i}(wc_{Bi}(z) - j_{Bi})w = 0. \quad (7)$$

Here the notation of the mass flow of the component $n = A, B$ is introduced

$$j_{ni} = c_{ni}(z)w_{ni}(z), \quad (8)$$

where $j_{ni} = const$ since the mass flow of any component does not depend on a coordinate in any section of the system. The solution to Eq. (7) is a sum of two exponents.

$$c_{Bi}(z) = A_{1Bi} \exp\left(\frac{wz}{D_i}\right) + A_{2Bi} \exp\left(-\frac{wz}{D_i}\right) + \frac{j_{Bi}}{w}. \quad (9)$$

The appearance of the new solution is natural when transforming systems of differential equations. However, as well as in the considered case, such solutions often do not satisfy an initial system of equations. In this case, solution (9) does not satisfy Fick's law (5). In the general case, Fick's law (5) and the diffusion equation are a part of the general system of mass, energy, and momentum transfer. The diffusion equation is an equation of the conservation of mass flow. That is, it is a part of the general system of transfer equations. It is reasonable to assume that the analysis of the process of solution phase transition by the equation of mass flow conservation only is incorrect. It is well-known that in the general case a pressure gradient arises in a multicomponent system due to a difference in component particle masses under isothermal diffusion. Under the action of this pressure gradient, forces affect the component particles. Let us introduce pressure into Fick's law (6). We will schematically derive generalized Fick's law [11] to formulate all the assumptions used in the statement of the diffusion problem with regard to pressure.

## 2. Introduction of pressure into Fick's law.

According to the Onsager relations, the diffusion flow of the component $n$ in a two-component homogeneous system is described by the equation

$$\mathbf{J}_n = L_{nA}\mathbf{X}_A + L_{nB}\mathbf{X}_B + L_{nu}\mathbf{X}_u, \quad (10)$$

here $\mathbf{J}_n$ is the diffusion flow of the component $n$, $L_{nA}, L_{nB}$ are the phenomenological coefficients, $\mathbf{X}_A, \mathbf{X}_B$ are the thermodynamic forces of a diffusion process, $\mathbf{X}_u$ is the thermodynamic force of a thermal conductivity process. The thermodynamic forces have the expressions

$$\mathbf{X}_n = -\frac{d}{dz}\left(\frac{\mu_n(p(z),T(z),c_n(z))}{T(z)}\right), \quad \mathbf{X}_u = -\frac{d}{dz}\left(\frac{1}{T(z)}\right), \quad (11)$$

here $\mu_n(p(z),T(z),c_n(z))$ is the chemical potential, $p(z)$ and $T(z)$ are the pressure and temperature, $c_n(z)$ is the concentration of the component $n$. According to the definition, the component concentrations are connected by the relation $c_A(z)+c_B(z)=1$. By substituting the thermodynamic forces of (11) to (10), we obtain the following expression for a two-component solution

$$J_n = -L_{nA}\frac{d}{dz}\left(\frac{\mu_A(p(z),T(z),c_A(z))}{T(z)}\right) - L_{nB}\frac{d}{dz}\left(\frac{\mu_B(p(z),T(z),c_B(z))}{T(z)}\right) - L_{nu}\frac{d}{dz}\left(\frac{1}{T(z)}\right). \quad (12)$$

The present work aims at analyzing a change in the component mass flows in homogeneous regions of a heterogeneous system as a result of a phase transition. During such analysis, difficulties arise in the physical interpretation of the condition of temperature constancy in phases. According to thermodynamics, the deviation of chemical potential from its equilibrium value at the interface is the motive force of the process of a phase transition. In experiments, the deviation of chemical potential from equilibrium is related to the specified velocity of interface motion through the field of a temperature gradient. In the theory of the growth of a solid phase from a single-component melt [3], the velocity of interface motion is a unique dependence on the kinetic undercooling $\Delta T_k$. Kinetic undercooling is the difference between the equilibrium temperature of the interface and its current temperature $\Delta T_k = T_e - T(0)$.

In the present work, the velocity of solution transfer is related to the partial velocities of the components. It is assumed that the partial velocities of the components depend on kinetic undercooling. However, in the isothermal problem $\Delta T_k = 0$. Therefore, we will assume that temperature gradients in phases are negligibly small to simplify the problem and consider it as the isothermal one. In this case, the isothermal problem is solved in each phase, but the temperature in the phases differs by $\Delta T_k$. On the one hand, this condition allows taking into account the deviation of the interface from equilibrium, i.e., taking into account the motive force of the process of a phase transition. On the other hand, it enables assuming temperature gradients to be zero in each phase. In this case, Eq. (12) takes the form

$$J_n = -L_{nA} \frac{d}{dz}\left(\frac{\mu_A(p(z),T,c_A(z))}{T}\right) - L_{nB} \frac{d}{dz}\left(\frac{\mu_B(p(z),T,c_B(z))}{T}\right).$$

We will not consider the cross-diffusion effects between the solution components in the problem under study to obtain the simplest solutions. Therefore, we assume that $L_{AB} = L_{BA} = 0$. In this case, component mass flows have the expressions

$$J_n = -L_{nn} \frac{d}{dz}\left(\frac{\mu_n(p(z),T,c_n(z))}{T}\right).$$

By differentiating, we find that

$$J_n = -L_{nn} \left[\frac{\partial \mu_n(p(z),T,c_n(z))}{T \partial c_n(z)} \frac{\partial c_n(z)}{\partial z} + \frac{\partial \mu_n(p(z),T,c_n(z))}{T \partial p(z)} \frac{\partial p(z)}{\partial z}\right].$$

The coefficients before derivative concentrations are diffusion coefficients [11]

$$D_A = \frac{L_{AA}}{T} \frac{\partial \mu_A(p(z),T,c_A(z))}{\partial c_A(z)}, \quad D_B = \frac{L_{BB}}{T} \frac{\partial \mu_B(p(z),T,c_B(z))}{\partial c_B(z)}.$$

The derivative chemical potential with respect to pressure at constant temperature and component concentration is the partial specific (molar) volume.

$$\vartheta_A = \frac{\partial \mu_A(p(z),T,c_A(z))}{T \partial p(z)}, \quad \vartheta_B = \frac{\partial \mu_B(p(z),T,c_B(z))}{T \partial p(z)}.$$

Specific values are used in the present work. As a result, we come to the system of equations

$$J_n = -D_n \frac{\partial c_n(z)}{\partial z} - \vartheta_{Ln} \frac{\partial p(z)}{\partial z}, \quad (13)$$

here the notations $\vartheta_{LA} = L_{AA}\vartheta_A$, $\vartheta_{LB} = L_{BB}\vartheta_B$ are introduced. In linear non-equilibrium thermodynamics [11], it is assumed that the coefficients of component diffusion are equal and the condition of such equality is found. However, the conditions of equality of diffusion coefficients for the isothermal, isobaric diffusion problem require the constancy of temperature and pressure. The condition of pressure constancy is not satisfied in the problem under consideration. We assume that in the problem under study the coefficients of component diffusion also are equal and find conditions under which this equality holds.

Let us equate the diffusion coefficients in system of equation (13)

$$\left(L_{AA}\vartheta_A + L_{BB}\vartheta_B\right)\frac{1}{T}\frac{\partial p(z)}{\partial z} + J_A + J_B = 0.$$

A sum of diffusion flows is zero, which follows from their definition. Consequently, the necessary condition of the equality of component diffusion coefficients is the equality

$$L_{AA}\vartheta_A = -L_{BB}\vartheta_B. \quad (14)$$

Let us find the relation between the diffusion coefficient and phenomenological coefficients. For this to be done, we use the Gibbs-Duhem equation. At constant temperature we obtain

$$c_A(z) \cdot d\mu_A(p(z), T, c_A(z)) + c_B(z) \cdot d\mu_B(p(z), T, c_B(z)) - V dp = 0.$$

We write the expressions of the differentials of chemical potentials as

$$d\mu_n(p(z), T, c_n(z)) = \frac{D_n T}{L_{nn}} dc_n + \vartheta_n dp(z),$$

then we substitute the relations $dc_A(z) = -dc_B(z)$ and $c_A(z) = 1 - c_B(z)$. As a result, we come to the equation

$$\left(\frac{D_B c_B(z)}{L_{BB}} - \frac{D_B(1-c_B(z))}{L_{AA}}\right) T dc_B + \left(\vartheta_B c_B(z) + (1-c_B(z))\vartheta_A - V\right) dp = 0.$$

The physical meaning of the Gibbs-Duhem equation is the relation between the intensive parameters of a solution at a small deviation from equilibrium. Consequently, in this equation volume is equal to the initial equilibrium unit volume $V_{\text{inf}}$. The value of chemical potential is equal to its value of an initial equilibrium solution, i.e., a solution at the concentration Cinf. Let us replace the notations and take into account relation (14)

$$\left(C_{inf} + \frac{(1-C_{inf})\vartheta_A}{\vartheta_B}\right)\frac{D_B T}{L_{BB}}dc_B + \left(\vartheta_B C_{inf} + (1-C_{inf})\vartheta_A - V_{inf}\right)dp = 0.$$

From here we find the relation between the diffusion coefficient and the phenomenological coefficient

$$L_{BB} = \frac{\left(C_{inf}\vartheta_B + (1-C_{inf})\vartheta_A\right)D_B T dc_B}{-\left(C_{inf}\vartheta_B + (1-C_{inf})\vartheta_A - V_{inf}\right)\vartheta_B dp}.$$

The volume depends on pressure and partial volume. We write out the volume differential

$$dV = \kappa dp + (\vartheta_B - \vartheta_A)dc_B.$$

Here $\kappa$ is the isothermal expansion coefficient, and the relation $dc_A(z) = -dc_B(z)$ is used. If we set a condition of volume constancy, $dV = 0$, and we obtain the relation between the potentials of concentration and pressure

$$dc_B = \frac{\kappa dp}{(\vartheta_A - \vartheta_B)}.$$

The expression for the phenomenological coefficient takes the form

$$L_{BB} = \frac{\left(C_{inf}\vartheta_B + (1-C_{inf})\vartheta_A\right)D_B T \kappa}{\left(C_{inf}\vartheta_B + (1-C_{inf})\vartheta_A - V_{inf}\right)(\vartheta_B - \vartheta_B)\vartheta_B}.$$

At equal diffusion coefficients $D_A = D_B = D$, initial generalized Fick's equations take the form

$$J_n = -D\frac{\partial c_n(z)}{\partial z} - \vartheta_{Ln}\frac{\partial p(z)}{\partial z}. \quad (15)$$

**3. Diffusion problem subject to pressure.**

Let us write Eq. (15) for the component $B$ of the phase $i$ in terms of coordinates fixed relative to the interface. We do not introduce the velocity of the interface ($w$ is the velocity of a solution) to the equations

$$c_{Bi}(z)(w_{Bi}(z) - w) = -D_i\frac{dc_{Bi}(z)}{dz} - \vartheta_{LBi}\frac{dp(z)}{dz}, \quad (16)$$

where $w_{Bi}(z)$ is the partial hydrodynamic velocity of the component $B_i$. This equation differs from Eq. (5) in an additional term. The additional term has simple physical meaning. It is a force that affects the particles of solution components due to a change in their partial velocities. Eqs. (16) and (6) constitute a system of equations relative to the concentration and velocity of the component $w_{Bi}(z)$. Formally, the unknown function $p(z)$ requires one more equation, an equation for the conservation of momentum. However, the obtained equations and the condition of one-dimensionality of the problem allow finding pressure related to a change in the momentum of component mass flows inside the heterogeneous system itself, i.e., pressure which arises due to a change in the momentum of the components of a solution under its interaction with the interface. Solution (9) obtained above satisfies the system of Eqs. (16) and (6) and the condition $j_{ni} = const$.

It is necessary to make a significant addition. Eq. (7) was obtained by the transformation of Fick's equation (5) and equation of mass flow conservation (6). If the system of Eqs. (16) and (6) is transformed in the same way, it is easy to obtain the same Eq. (7). Moreover, in special cases two more general solutions are obtained.

$$c_{Bi}(z) = A_{1Bi} \exp\left(\frac{wz}{2D_i}\right) \sin\left(\frac{\sqrt{3}wz}{2D_i}\right) + A_{2Bi} \exp\left(-\frac{wz}{2D_i}\right) \cos\left(\frac{\sqrt{3}wz}{2D_i}\right) + \frac{j_{Bi}}{w},$$

$$c_{Bi}(z) = A_{1Bi} \sin\left(\frac{wz}{D_i}\right) + A_{2Bi} \cos\left(-\frac{wz}{D_i}\right) + \frac{j_{Bi}}{w}.$$

If one knows the distribution of the concentration $c_{Bi}(z)$, it is easy to find pressure distribution from Eq. (16)

$$\vartheta_{LBi} p(z) = \int \left( c_{Bi}(z) w - D_i \frac{\partial c_{Bi}(z)}{\partial z} - j_{Bi} \right) dz + A_{pBi}. \quad (17)$$

The integration constant $A_{pBi}$ does not depend on $z$. The velocity of any component $n$ is determined from the expression $w_{ni}(z) = j_{ni} / c_{ni}(z)$. Solution (9) is considered as a general solution in the present work. At $\vartheta_{LBi} = 0$, solution (9) does not satisfy Fick's law (16). For this reason, the solution to the quasi-equilibrium problem [3] gives constant concentration in one of the phases (4). The substitution of general solution (9) to (17) gives pressure distribution in phases.

$$p_{sol}(z) = -\frac{2A_{2sol}D_{sol}}{\vartheta_{LBsol}} \exp\left(-\frac{wz}{D_{sol}}\right) + PS(0) \quad (18)$$

$$p_{liq}(z) = -\frac{2A_{2liq}D_{liq}}{\vartheta_{LBliq}} \exp\left(-\frac{wz}{D_{liq}}\right) + PL(0) \quad (19)$$

**4. Solution to the boundary diffusion problem.**

The momentum of a component changes when its velocity changes. The difference in component momentum on the opposite sites of the interface results in the appearance of an additional force affecting the component. The forces arising between components under phase transitions are the internal forces of a system. These forces can manifest themselves only if the system is affected by external forces which move the interface. There may be a variety of reasons for the motion of the interface. For example, this is the motion of a temperature gradient through a solution or phase transition in a metastable solution. In each individual case, the physical scheme of external phenomena leading to a phase transition may be different. In the problem under consideration, it is assumed that a solution moves through a stationary temperature field. This scheme is usually used to produce materials using phase transitions. It is obvious that the statement of the problem under study is simplified. Instead of a force moving a solution, the velocity of solution transfer is set here. This simplification is natural for a stationary problem. However, the thermodynamic properties of a solution, for example, pressure, change when applying a force to it. The pressure is transmitted to the interface. The interface shifts and chemical potential deviates from its equilibrium value at it. As a result, it is the deviation of chemical potential from equilibrium that is the motive force of a phase transition and all the phenomena which accompany the phase transition. The term containing pressure in Eq. (16) is a force affecting the component $B$. However, this term is related to the pressure only which arises as a result of component diffusion. Below, when interpreting the solutions to the problem, it will be necessary to introduce pressure related to external action. The present work does not consider the general statement of the problem with an external force. This statement unreasonably complicates the problem. The solution under study aims at demonstrating one of the reasons for the segregation of the components of a solution under its phase transition.

Using generalized Fick's law (16) and solution (9), after transformations we come to the equation

$$2w(A_{2Bsol} - A_{2Bliq}) + j_{Bsol} - j_{Bliq} = 0. \quad (20)$$

Mass flows were included in the boundary condition again after the substitution of solutions. Under condition (20) the diffusion problem in the infinite intervals of phases has a final solution. However, in this case, the mass flows of components in phases have different values. Two solutions only have a physical meaning at which the condition for the equality of phase mass flows $j_{Bsol} - j_{Bliq} = 0$ is fulfilled. The pressures in phases in these solutions are constant. They formally agree with the quasi-equilibrium solutions from [3]. Consequently, condition (20) is the condition for the conservation of the momentum of a solution component. Under the condition of the conservation of a component mass flow $j_{Bsol} - j_{Bliq} = 0$, (20) gives the relation between the integration constants $A_{2Bsol} = A_{2Bliq}$. The constant $A_{2Bliq}$ is a coefficient at the divergent exponent of the solution for the liquid phase. If one considers the liquid phase in a semi-infinite interval, i.e., if one assumes $A_{2Bliq} = 0$, $A_{2Bsol} = 0$. We come to the problem in the infinite intervals of phases. Therefore, we assume that the liquid phase is limited by the $z_0$ coordinate with the specified concentration $c_{Bliq}(z_0)$. From Eqs. (20), $c_{Bliq}(z_0) = C_{ini}$, and $j_{Bsol} - j_{Bliq} = 0$, $A_{2Bliq}$, $A_{2Bsol}$, and $A_{1Bliq}$ are found. If one substitutes them to the general solution, it will depend on the values of concentration and component velocities at the interface. In total, there are eight unknown quantities.

The values of concentrations and partial velocities at any point of the system are connected by the relations

$$c_{Ai}(z) + c_{Bi}(z) = 1 \quad w = c_{Ai}(z)w_{Ai}(z) + c_{Bi}(z)w_{Bi}(z). \quad (21)$$

In addition to these four equations, solutions provide the relations between concentrations and velocities at the interface. In total, we have six equations.

One more equation is the relation between boundary concentrations and a phase diagram. As the simplest example of an equilibrium phase diagram, we consider a eutectic phase diagram with linear equilibrium lines. The present work takes into account the deviation of chemical potential from its equilibrium value at the interface. Consequently, within the temperature-concentration diagram under a phase transition, the temperature of the interface will differ from the equilibrium value by the value of the kinetic undercooling $\Delta T_k$. The interface is assumed to be sharp, i.e., the boundary temperature of the liquid phase is equal to the boundary temperature of the solid phase. Under this condition, the relation between boundary phase concentrations in the interval of concentration values between 0 and the eutectic concentration $c_{eut}$ is expressed through the constant segregation coefficient $k_e = c_{Bsol}(0)/c_{Bliq}(0)$. The obtained system of seven equations contains eight unknown concentrations and partial velocities at the interface. Let us express seven concentrations and partial velocities through the partial velocity $w_{Bliq}(0)$. As a result of the calculations and transformations, the solutions to the problems can be written as

$$c_{Bsol}(z) = \frac{c_{Bliq}(z_0)}{a}\left((k_e w - w_{Bliq}(0))\exp\left(-\frac{wz}{D_{sol}}\right) + w_{Bliq}(0)\right) \quad (22)$$

$$c_{Bliq}(z) = \frac{w_{Bliq}(0)}{ab}\left(\left(\exp\left(\frac{w(z-z_0)}{D_{liq}}\right) + \exp\left(\frac{wz_0}{D_{liq}}\right) - \exp\left(-\frac{w(z-z_0)}{D_{liq}}\right)\right)(1-k_e) + k_e \exp\left(-\frac{wz_0}{D_{liq}}\right)\right) + \\ \frac{w_{Bliq}(0)}{ab}\left(-(a + w_{Bliq}(0))\left((1-k_e)\exp\left(\frac{wz}{D_{liq}}\right) + k_e \exp\left(-\frac{wz}{D_{liq}}\right)\right)\right) \quad (23)$$

$$p_{sol}(z) = \frac{2AD_{sol}}{\vartheta_{LBsol}}\left(1 - \exp\left(-\frac{wz}{D_{sol}}\right)\right) + PS(0) \quad (24)$$

$$p_{liq}(z) = \frac{2AD_{liq}}{\vartheta_{LBliq}}\left(1 - \exp\left(-\frac{wz}{D_{liq}}\right)\right) + PL(0) \quad (25)$$

here

$$a = w(k_e - 1)\exp\left(\frac{wz_0}{D_{liq}}\right) + (w_{Bliq}(0) - k_e w)\exp\left(-\frac{wz_0}{D_{liq}}\right) - w_{Bliq}(0)$$

$$b = (k_e - 1)\exp\left(\frac{wz_0}{D_{liq}}\right) - k_e \exp\left(-\frac{wz_0}{D_{liq}}\right)$$

$$A = -\frac{k_e c_{Bliq}(z_0)}{b}\left(1 - \frac{w_{Bliq}(0)}{a}\left((k_e - 1)\exp\left(\frac{wz_0}{D_{liq}}\right) - 1\right)\right)$$

**5. Kinetics of the addition of the particles of liquid solution components to a new phase.**

The solution velocity and partial velocity of the component $B$ of the liquid phase were included in solutions (22) – (25). In the limit of system transition to equilibrium, these velocities should vanish. However, if, for example, one substitutes $w=0$ to the solution, the solution gives concentration divergence in both phases. To obtain equilibrium of the system, we note that the velocities are included in the system solutions, excepting the exponent of the exponential, in the form of the relation $\gamma = w/w_{Bliq}(0)$. If $\gamma \to 1$ at $w_{Bliq}(0) \to 0$ and $w \to 0$, we obtain the system equilibrium $c_{Bliq}(z) = const$, $c_{Bsol}(z) = const$, and $c_{Bsol}(z)/c_{Bliq}(z) = k_e$. The limiting transition to equilibrium means vanishing of the motive force of a phase transition, i.e., the deviation of chemical potential at the interface vanishes. It is well-known that the crystallization rate of a single-component melt is determined by the kinetics of the addition of melt particle components to a new growing phase [3]. Therefore, it is reasonable to assume that the velocity of melt components is determined by the kinetics of the addition of the particles of a specific component to the solid phase. The solution velocity $w$ depends on the partial component velocities by relation (21). Let us consider the dependence of the velocity $w_{Bliq}(0)$ on the deviation of the chemical potential of this component from equilibrium at the interface.

$$w_{Bliq}(0) = W\left(\Delta\mu_{Bliq}\left(\Delta T_k, \Delta p, \Delta c_{Bliq}\right)\right). \quad (26)$$

In this expression

$$\Delta\mu_{Bliq}\left(\Delta T_k, \Delta p, \Delta c_{Bliq}\right) = \mu_{Bliq}\left(T_e, p_e, c_{eBliq}\right) - \mu_{Bliq}\left(T(0), p(0), c_{Bliq}(0)\right),$$

$$\Delta T_k = T_e - T(0), \quad \Delta p = p_e - p(0), \quad \Delta c_{Bliq} = c_{eBliq} - c_{Bliq}(0).$$

We assume the deviation of chemical potential from equilibrium at the interface to be small. A small deviation of temperature and concentration from equilibrium corresponds to it. It is well-known from the experiments that a change in the pressure under normal experimental conditions has little effect on the equilibrium values of chemical potential. Therefore, the temperature-concentration equilibrium diagram is usually used in calculations. As stated above, this phase diagram is used in the present work. That is, a change in the chemical potential due to pressure at the interface is not taken into account when calculating the velocity. Therefore, Eq. (26) can be written in the form

$$w_{Bliq}(0) = W\left(\Delta\mu_{Bliq}\left(\Delta T, \Delta c_{Bliq}\right)\right) \approx W\left(\mu_{Bliq}\left(T_e, c_{eBliq}\right)\right) + \frac{\partial\mu_{Bliq}}{\partial T}\Delta T_k + \frac{\partial\mu_{Bliq}}{\partial c_{Bliq}}\Delta c_{Bliq}. \quad (27)$$

The equilibrium phase diagram represents dependencies between the equilibrium values of temperature and concentration $T_e = S(c_{eBsol})$ and $T_e = L(c_{eBliq})$. At small deviations of the temperature and concentration of the liquid phase from equilibrium their increments are connected by the linear dependence

$$T(0) \approx T_e + \left.\frac{\partial L}{\partial c_{Bliq}}\right|_{\substack{c_{Bliq}=c_{eBliq}\\T=T_e}} \Delta c_{Bliq}.$$

From here we obtain the relation between the increments of concentration and temperature

$$\Delta c_{Bliq} = -\left(\left.\frac{\partial L}{\partial c_{Bliq}}\right|_{\substack{c_{Bliq}=c_{eBliq}\\T=T_e}}\right)^{-1} \Delta T_k.$$

We substitute this expression to Eq. (27)

$$w_{Bliq}(0) \approx \left( \frac{\partial \mu_{Bliq}}{\partial T} - \frac{\partial \mu_{Bliq}}{\partial c_{Bliq}} \left( \frac{\partial L}{\partial c_{Bliq}} \bigg|_{\substack{c_{Bliq}=c_{eBliq} \\ T=T_e}} \right) \right)^{-1} \Delta T_k = h_\mu \Delta T_k.$$

Here the condition $W(\mu_{Bliq}(T_e, c_{eBliq})) = 0$ is used and the notation of the kinetic coefficient, which is assumed to be known, is introduced. The obtained relations between $\Delta T_k$, $w_{Bliq}(0)$ and $w$ show that these dependencies have the following properties.

1. The functions $w$ and $w_{Bliq}(0)$ are continuous dependencies on the kinetic undercooling $\Delta T_k$.

2. The functions $w$ and $w_{Bliq}(0)$ vanish in the limit $\Delta T_k \to 0$.

3. $\lim\limits_{\Delta T_k \to 0} \gamma = 1$.

These properties are enough to draw some conclusions on the expansion of the functions $w_{Bliq}(0)$ and $w$ into the Maclaurin series with respect to a small deviation of the system from equilibrium, i.e., with respect to $\Delta T_k$.

1. Expansion of the functions $w_{liq}(0)$ and $w$ into the Maclaurin series with respect to the kinetic undercooling $\Delta T_k$ does not contain a zero-order term.

2. Terms with the smallest powers of this expansion have the same powers.

3. Coefficients at the smallest powers of $\Delta T_k$ are equal.

In the simplest case, the dependencies of $w_{Bliq}(0)$ and $w$ on $\Delta T_k$, which satisfy these three conditions, can be formed from the functions $\Delta T_k$ and $\Delta T_k^2$. If one requires that solutions (22), (23) describe the system at arbitrary values of the segregation coefficient, one should take into account the dependencies of $w_{Bliq}(0)$ and $w$ on $k_e$. In this case, dependencies (22) and (23) provide an additional condition. In the special case of azeotropic solutions, i.e., at $k_e = 1$, the composition of both phases should be the same. For this to be done, the satisfaction of the equality $w_{Bliq}(0) = w$ at $k_e = 1$ is enough. The simplest dependencies of $w_{Bliq}(0)$ and $w$, which satisfy these three conditions, are the functions

$$w = k_e h \Delta T_k + (1 - k_e) h_B \Delta T_k^2, \quad w_{Bliq}(0) = k_e h \Delta T_k, \quad (28)$$

where $h$ and $h_B$ are the kinetic coefficients. The equality of the first coefficients determines the condition of system equilibrium. By substituting (28) to solutions (22) – (25), we obtain the dependencies of pressure and concentration distribution in phases on the spatial variable z and kinetic undercooling $\Delta T_k$. These functions are simply obtained and are cumbersome, that is why they are not written out here. We denote them by $c_{Bsol}(z, \Delta T_k)$, $c_{Bliq}(z, \Delta T_k)$, $p_{Bsol}(z, \Delta T_k)$, and $p_{Bliq}(z, \Delta T_k)$ according to expressions (22), (23). The physical limitation of concentrations is limitations related to the phase diagram.

This fragment is crucial to understanding the processes of the mass transfer of solutions through the interface. It is the kinetics of the addition of component particles to a new growing phase that provides a change in the component momentum under phase transitions. The introduction of kinetic dependencies (28) in the problem under study is formal. In reality, kinetic dependencies should be determined for the velocities $w_{Aliq}(z)$ and $w_{Bliq}(z)$. The solution velocity is connected to these velocities by relation (21). Pressure distribution in phases (24), (25) contains two unknown quantities $PS(0)$ and $PL(0)$. In the general case, these parameters depend on $\Delta T_k$ and are found from boundary conditions. The pressures $PS(0)$ and $PL(0)$ depend on the external conditions under which a phase transition occurs. As an example of the physical interpretation of the problem, we shall assume that the reason for a phase transition is pulling of the solid phase from a melt. Let the pulling force be proportional to the solution velocity. Without going into the details of the pulling method, we set pressure (pulling force) on the side of the solid phase in the form $PS(0) = H\Delta T_k + P_{ext}$. We assume that on the side of a liquid solution and throughout the liquid phase the pressure is equal to the external pressure $PL(0) = P_{ext}$.

### 6. Model calculation of the distribution of concentration and pressure.

Let us consider the results of the numerical calculation of concentration and pressure distribution in the system the parameters of which are shown in Table 1. The values of calculation parameters are selected so that it is convenient to interpret the physical relation between velocities, concentration, and pressure in the graphs. The concentration dependencies on spatial coordinate and kinetic undercooling are presented in Fig. 1. On the right, there is concentration distribution in liquid phase 1, on the left there is that in solid phase 2.

| $k_e$ | 0.6 | $C_{ini}$ | 0,15 |
|---|---|---|---|
| $H$ | 0 | $h$ | -0.1 |
| $P_{ext}$ | 1 | $h_B$ | -2.8 |
| $\vartheta_{LBliq}$ | -4.5 | $\vartheta_{LBsol}$ | 0.1 |
| $D_{liq}$ | 1 | $D_{sol}$ | 0.1 |

Table 1. Values of system parameters for numerical calculations.

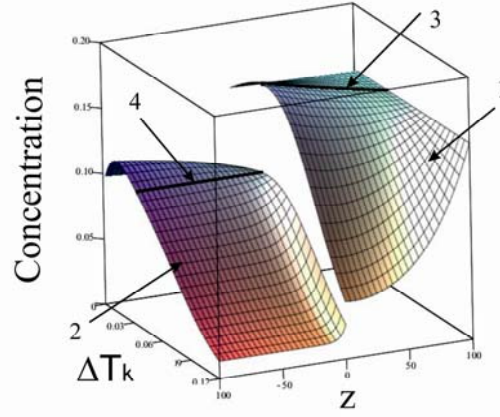

Figure 1. Concentration distribution in the vicinity of the interface. 1 – Liquid phase; 2 – solid phase; 3, 4 – concentration distribution at $\Delta T_k = \Delta T_Q$.

Concentration distribution changes continuously throughout the region of phase existence of the region of equilibrium phase diagram under consideration $0 < C < C_{eut}$. The concentration changes exponentially in both the liquid and solid phases. Curves 3 and 4 emphasize concentration distribution in which the concentration of the solid phase is constant. From solution (22) it follows that the relation $w_{Bliq}(0) = k_e w$, i.e., the equality of the coefficient to zero at the exponent of the distribution of solid phase concentration, corresponds to this regime. In this form, the obtained solution formally agrees with quasi-equilibrium solution (4) from [3]. However, the physical meaning of these solutions is different. Solution (4) represents the dependence of concentration distribution in solution phases at different values of the solution velocity $w$. As follows from the solution, when the solution velocity changes, the value of phase concentration at the interface remains constant at any values of $w$. This result is due to the fact that the quasi-equilibrium statement of the problem does not take into account the relations between partial velocities of solution components and deviation of the solution from equilibrium. From Eq. $w_{Bliq}(0) = k_e w$ it is easy to find the value of kinetic undercooling for a regime denoted by curves 3 and 4

$$\Delta T_Q = h / h_b.$$

The solution at $\Delta T_k = \Delta T_Q$ separates concentration distribution into two regions. At $\Delta T_k < \Delta T_Q$, the distribution of solid phase concentration decreases from the interface to the infinite point $z = -\infty$. At $\Delta T_k > \Delta T_Q$, the distribution of solid phase concentration increases. Let us consider the reason for such a change in the concentration.

Pressure distribution in liquid phase 1 and solid phase 2 is shown in Fig. 2.

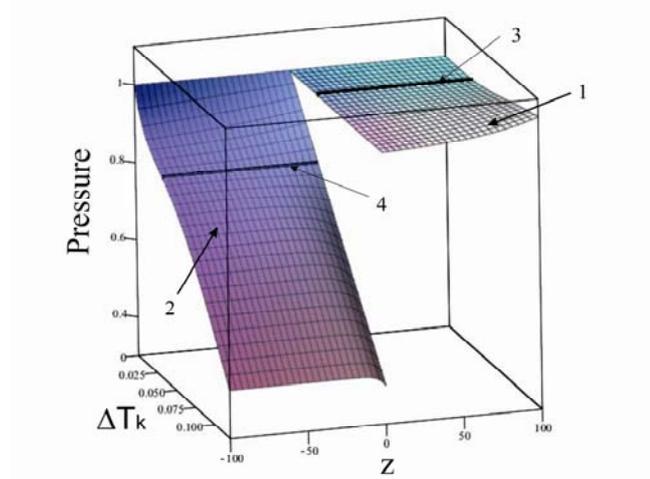

Figure 2. Pressure distribution in the vicinity of the interface. 1 – Liquid phase; 2 – solid phase; 3, 4 – pressure distribution at $\Delta T_k = \Delta T_Q$.

In the solid phase, the pressure consists of the pressure of external action and the pressure related to diffusion. In the graph, one cannot see a change in the pressure due to component diffusion in the solid phase. The reason is the selected value of the external action *H*. To set *H* in solution (24), pressure distribution was plotted preliminarily not taking into account the external action, *H*=0. Then, *H* was selected so that $p(z, \Delta T_k, c_B)$ was less than external pressure. To show pressure arising due to diffusion, Fig. 3 presents separately the dependence of pressure distribution in the solid phase at *H*=0. The pressure distribution is illustrated in liquid phase 1 and solid phase 2. Curve 3 shows the pressure which corresponds to the regime $\Delta T_k = \Delta T_Q$ in which concentration distribution in the solid phase is constant (to concentration distribution 3, 4 in Fig. 1). Eq. (24) provides the same condition $w_{Bliq}(0) = k_e w$ for the regime of constant pressure in the solid phase as the condition for constant concentration. Fig. 3 presents that in the interval $0 < \Delta T_k < \Delta T_Q$ pressure decreases, and in the interval $\Delta T_k > \Delta T_Q$ it increases when the *z* coordinate changes from -∞ to 0. Note that for illustrative purposes the axes in Fig. 3 are shifted with respect to the axes in Fig. 1.

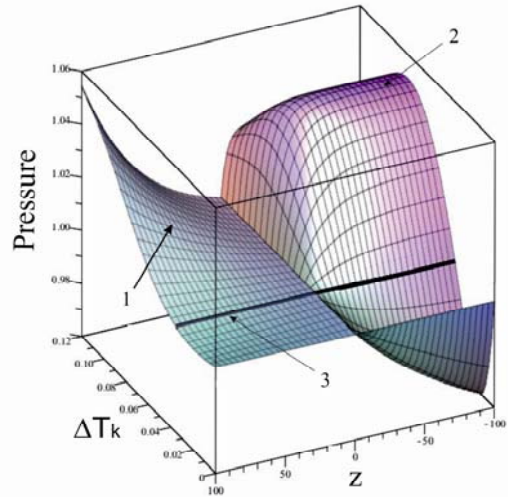

Figure 3. Pressure distribution in the vicinity of the interface at $H$=0. 1 – Liquid phase; 2 – solid phase; 3 – pressure distribution at $\Delta T_k = \Delta T_Q$.

The right side of Eq. (16) is forces affecting the corresponding solution components. A concentration gradient is a force arising due to the non-equilibrium distribution of a component. A pressure gradient is a force arising due to the non-equilibrium distribution of momentum between solution components. This force arises due to the interaction between solution components and the interface. The directions of the forces depend on the direction of the corresponding gradients, signs of diffusion coefficients, and partial component volume. According to solution (24) (25), if partial volume changes its sign, the direction of a pressure gradient changes to the opposite one. Under condition (14), the forces affecting components are equal in value and opposite in sigh that is why they become equal. The equality of these forces agrees completely with Newton's third law. The equality of forces in the liquid phase means that a liquid moves as a single whole without hydrodynamic flows.

At positive value $\vartheta_{LBsol}$, in the interval $0 < \Delta T_k < \Delta T_Q$ the particles of the component «B» are affected by forces directed away from the interface. Therefore, component concentration increases away from the interface. In the interval $\Delta T_k > \Delta T_Q$, there is a reverse situation. In this case, pressure decreases towards the interface. Consequently, the particles of the component «B» are affected by a force directed towards the interface. As a result, concentration increases in the direction towards the interface.

### 7. Discussion and conclusions.

The description of a stationary phase transition of solutions, provided in the present work, apparently is the simplest model of phase transitions. The main conclusion from the considered problem is that even in the simplest case the problem of component distribution during the phase transition of a solution should contain all the variables of chemical potential – temperature, pressure, and density of components. Indeed, if we do not take into account a change in the temperature at the interface, we come to the quasi-equilibrium solution. In this form, the solutions do not provide the limiting transition to equilibrium since the relation between the velocities and the kinetics of the addition of component particles to a new phase is lost. If the pressure is not taken into account, the problem provides a solution only which formally agrees with the solution to quasi-equilibrium problem (4). This solution corresponds to the value $\Delta T_k = \Delta T_Q$, i.e., trajectories 3 and 4 in Fig. 1 and trajectory 3 in Fig. 3.

The quasi-equilibrium solution is used, for example, in the Burton–Primm–Schlicter (BPS) theory of zone refining of solutions under phase transitions [3]. An effective segregation coefficient is introduced in the BPS theory to calculate a segregation degree. This coefficient is equal to the ratio of the specified concentration of a liquid solution at the point $z_0$ to the boundary concentration of a solid solution. Curves 5, 6, and 7 in Fig. 4 show the dependencies of the effective segregation coefficient of the BPS theory on the solution velocity w. Curves 2, 3, and 4 in Fig. 4 illustrate similar dependencies of the problem considered here. At $z_0 = 0$, the coefficients in both cases are equal to $k_e$ (Curve 1 in Fig. 4). The dependencies differ qualitatively. The explanation of this is obvious. In the proposed solution, an additional degree of freedom arises, i.e., pressure which affects the segregation of solution components by the interface.

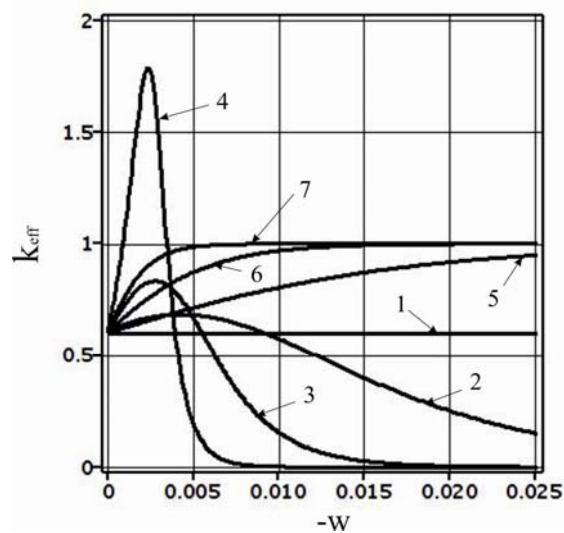

Figure 4. Dependence of the effective distribution coefficient on the solution velocity. 1, 2, 3, 4 – the solution obtained in the article; 1, 5, 6, 7 - according to the BPS theory. 1. - $z_0 = 0$; 2, 5 - $z_0 = 100$; 3, 6 - $z_0 = 300$; 4, 7 - $z_0 = 750$.

As stated above, the present work is motivated by component segregation under a phase transition, which was observed in the experiments [9,10]. Let us consider the result of component segregation of solutions (22) and (23). Concentration distributions in phases (22) and (23) in the equilibrium $\Delta T_k = 0$ provide the equilibrium values of the phase concentrations $c_{Bliq}|_{\Delta T_k=0} = C_{ini}$, $c_{Bsol}|_{\Delta T_k=0} = k_e C_{ini}$. The concentration at the infinite point of the solid phase depends on the value of kinetic undercooling, i.e., on the velocity of solution transfer. However, the value of the limit of the function $c_{Bsol}(z)$ at $z \to \infty$ and $\Delta T \to 0$ depends on the sequence of the limiting transition:

$$\lim_{z \to \infty} \exp\left(-\frac{(k_e h + (1-k_e) h_B \Delta T_k) \Delta T_k z}{D_{sol}}\right) = 0, \quad \lim_{\Delta T_k \to 0} \exp\left(-\frac{(k_e h + (1-k_e) h_B \Delta T_k) \Delta T_k z}{D_{sol}}\right) = 1.$$

This is an unavoidable drawback of the problem solution in infinite intervals. To avoid this uncertainty, one should take into account real physical situation and solve the problem of concentration distribution by setting boundary conditions at the finite values of spatial coordinate. Within the obtained solutions, Fig. 5 demonstrates the dependence of the concentration $c_{Bsol}(z)$ on kinetic undercooling for several values of $z$.

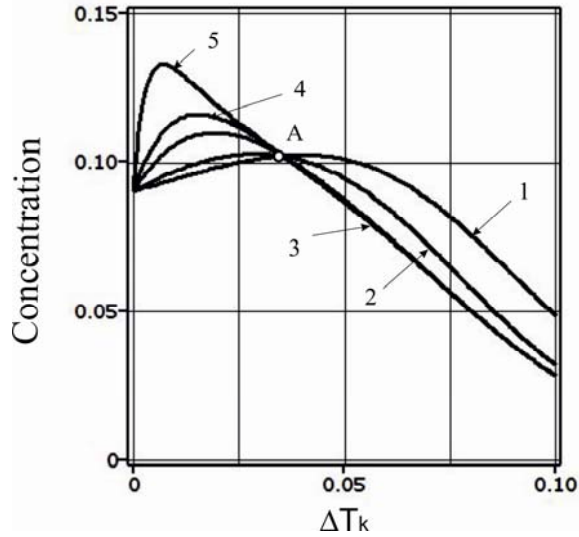

Figure 5. Concentration dependence of the component $B$ on kinetic undercooling for different values of $z$. 1. $z=0$; 1. $z=-10$; 2. $z=-50$; 3. $z=-100$; 4. $z=-500$.

The point of intersection of all the solutions $A$ is the solution at $\Delta T_k = \Delta T_Q$ indicated with 4 in Fig. 1. It is obvious that not all the obtained solutions can be observed in experiments. It is necessary to note two reasons for which the considered model is a rough approximation to real phase transitions. First, this is severe restrictions imposed during the construction of the model.

Second, as is well-known, it is difficult to obtain a stable plane interface in experiments. Therefore, the question arises on the stability boundary of the obtained solutions.

The results of the present work change significantly the description of the processes of mass transfer at the interface. The kinetics of the addition of solution components to a new phase leads to a change in the momentum of components under their transfer through the interface. Additional pressure arises in phases as a result of a change in the momentum. This additional pressure crucially affects the distribution of component concentration.

**Reference.**